# The Asymptotic Behavior of Minimum Buffer Size Requirements in Large P2P Streaming Networks

Lei Ying
Dept. of ECE
Iowa State University
Email: leiying@iastate.edu

R. Srikant
Dept. of ECE and CSL
University of Illinois at Urbana-Champaign
Email: rsrikant@illinois.edu

Srinivas Shakkottai
Dept. of ECE
Texas A&M University
Email: sshakkot@tamu.edu

*Abstract*—The growth of real-time content streaming over the Internet has resulted in the use of peer-to-peer (P2P) approaches for scalable content delivery. In such P2P streaming systems, each peer maintains a playout buffer of content chunks which it attempts to fill by contacting other peers in the network. The objective is to ensure that the chunk to be played out is available with high probability while keeping the buffer size small. Given that a particular peer has been selected, a *policy* is a rule that suggests which chunks should be requested by the peer from other peers.. We consider consider a number of recently suggested policies consistent with buffer minimization for a given target of skip free playout. We first study a *rarest-first* policy that attempts to obtain chunks farthest from playout, and a *greedy* policy that attempts to obtain chunks nearest to playout. We show that they both have similar buffer scalings (as a function of the number of peers of target probability of skip-free probability). We then study a hybrid policy which achieves order sense improvements over both policies and can achieve order optimal performance. We validate our results using simulations.

## I. INTRODUCTION

Peer-to-peer (P2P) networks are rapidly becoming the medium of choice for content distribution over the Internet. Several studies have indicated their widespread adoption, and suggest that anywhere between 35-90% of Internet bandwidth is consumed by P2P applications [1], [2]. The initial application of the P2P idea was for file sharing, but as P2P file sharing networks have matured, many ideas have been transplanted to media streaming applications with a degree of success (*eg.* [3], [4]). As entertainment delivery over the Internet becomes increasingly main stream, P2P streaming is likely to assume an even greater significance.

Streaming media using P2P has a more exacting set of constraints than file sharing since each *chunk* that constitutes the streamed file must be received within a certain deadline in order to allow for smooth sequential playing out of the media. A natural approach for designing streaming networks that inherently possesses this sequential nature of streaming is to create a multicast tree amongst the source and users either at the IP or higher layer [5]–[7], and to *push* chunks on the tree. However, this approach often needs significant infrastructural overheads to perform adequately; for example, an IP layer multicast requires support by the routers in the network. Another issue is that when end-users act as the peer nodes in the multicast tree, the churn caused by users arriving and departing causes significant inefficiencies [8].

A more popular approach to P2P Streaming is using application layer multicast with a full mesh topology among peers. The idea is to maintain a playout buffer, and to *pull* chunks into the buffer by communicating with a random selection of peers. This approach bears a strong resemblance to the BitTorrent technique [9] of selecting peers from a full mesh in order to exchange chunks [9]. However, the sequential playout constraint requires that the chunk to be played next is available when required. Several systems exist that use this approach to content streaming, and notable examples are CoolStreaming [10] that was one of the first systems to employ this approach, and some that are highly popular in East Asia such as PPLive [3], QQLive [4] and TVAnts [11].

In this paper we study optimal algorithms for mesh-based P2P multicast. Here, a server generates chunks at constant rate, and randomly selects on peer to push it to. Other peers select some other peer at random, and request some chunk in the selected peer's possession. The objective is to ensure that the chunk that needs to be played out is available with high probability. Under this paradigm, a distribution algorithm essentially answers the following question: *Given that a particular peer has been selected, which chunk from that peer should be pulled'?* The answer is not obvious, since on the one hand pulling the chunk with the shortest playout deadline might be optimal in the short term, but might yield poor performance as far as the rest of the peers are concerned. On the other hand, pulling a chunk that is farthest from playout would give maximum amount of time for its dispersal amongst peers, but this might be at the expense of the chunk that is most urgently required. It is also clear that the question cannot be answered independently of the buffer size, since a large buffer size gives more time for chunk dispersal than a short one.

There has been considerable interest in P2P systems for both file delivery and streaming. Analytical models and studies of the achievable limits of P2P file delivery are presented in [12]–[16]. The primary objective of these studies is to ensure that all interested peers obtain the entire file with as short a delay as possible. Our current work uses some of the techniques used in these papers, but since we focus live streaming, the objective is somewhat different due to the sequential nature of playout. There have also been analytical and simulation studies of tree and mesh based P2P streaming [8], [17]. In [8], mesh versus tree approaches are considered primarily using simulations, whereas in [17] upper bounds on performance are developed. Tree-based approaches are considered in [5]–[7], but since we consider end-user P2P, we focus on a full mesh pull based approach. The study in [18] considers using BitTorrent to directly assist in live streaming. Finally, the authors of [19], [20] consider a general class of streaming algorithms. Their main contribution is to develop an analytical model of P2P real-time streaming algorithms, and numerically show that based on available server capacity, different hybrids of greedy and rarest-first policies perform better than either of the constituent policies. Our work develops on the model introduced in [19], but our goal is to analytically answer the following question: how much buffer size reduction can be achieved using a hybrid policy? We are interested in this question since a large buffer size increases start-up latency and adversely impacts the "real-time" nature of the system. The model in [20] is slightly more accurate than the one in [19], but the model in [19] is more tractable which allows us to obtain design insights which are confirmed by our simulation results.

*Main results*

The system that we consider consists of $M$ users who are all simultaneously interested in a real-time content stream generated by a server. The stream consists of chunks with one new chunk generated at each discrete time instant, and the server selects one peer at random for each new chunk. Peers obtain chunks either if they are selected by the server, or by full-mesh P2P with random peer selection. Peers maintain a buffer of size $m$, with the chunk in the $m$th location played out at each time instant if available. We have a target of skip-free playout probability of $q$ over all peers.

We present an analytical characterization of the scaling of required buffer size $m$ with $M$ and $q$ under different policies. The main results of of our analysis are summarized below:

- We consider the *rarest-first policy* wherein priority is given to the chunks farthest from playout. In other words, a peer picks that chunk in the difference set with his selected peer that is closest to the first buffer position. We show that given any target probability $0.5 < q \leq 1$, the buffer size to attain this target probability with the rarest-first policy scales approximately as

$$\log(M) + \log(2q - 1) + \frac{1}{2(1-q)}.$$

- We then consider the *greedy policy* wherein priority is given to the chunks closest to playout. Thus, a peer picks that chunk in the difference set with his selected peer that is closest to the $m$th buffer position. We show that given any target probability $q \geq \frac{1}{M}$, the buffer size to attain this target probability with the greedy policy approximately scales in a similar fashion as

$$\log(M) + \log(q) + \frac{1}{1 - q + \frac{2}{M}}.$$

- We develop a *hybrid policy* $h_\epsilon$ which combines the greedy and the rarest-first policies. The policy uses rarest-first up to a buffer position where the probability of occupancy is greater than $\epsilon$, and switches to using the greedy policy from there on. For this policy, we show that there exist constants $a_\epsilon$ and $a_{\epsilon,\epsilon_2}$, independent of $M$ and $q$, such that, if the buffer size $m$ satisfies

$$m \geq a \log(M) + a_{\epsilon,\epsilon_2} \log\left(\frac{1}{1-q}\right),$$

then the skip-free playout probability $p_{(h_\epsilon,m)}(m) \geq q$. Thus, the buffer size required by this policy less than that required by either of the two policies it is composed of in an order sense.

Above, $\log(.)$ refers to base 2. Consider an example scenario where we have $M = 10,000$ and a target probability of skip-free playout $q = 99.9\%$, we see that both the rarest-first and greedy policies would need a buffer size of the order of 1000s, whereas the hybrid policy would need a buffer size only of the order of 10s. We contrast the results with a straightforward lower bound. Given a target probability $\frac{1}{M} \leq q \leq 1$, the buffer size to attain this target probability under any policy should be at least $\log M$, which for our example is also of the order of 10s. Further, for a target of $q \geq 1 - 1/M$, the hybrid policy attains this lower bound in an order sense.

Finally, we show using simulations that the order sense reduction in buffer size promised by the hybrid-policy do indeed materialize, particularly when the desired target of buffer occupancy is high, so arguing for the adoption of such hybrid chunk selection policies in P2P streaming systems.

## II. BACKGROUND AND MODEL

We first introduce the model of a P2P content streaming system developed in [19] under a large-system assumption. According to the model, time is considered to be slotted and is synchronized across the whole system. The system is illustrated at some time slot in Figure 1, and consists of a single server and $M$ peers. The server is a source of real-time content such as media coverage of a live event, and generates one chunk of new data per time slot. The server has a limited communication capacity, and is able to transmit this chunk to exactly one peer. It does so by choosing one of the $M$ peers at random. Each peer is assumed to have a buffer of size $m$, with the buffer positions indexed by $i \in \{1,..,m\}$. As shown in Figure 1, in each time slot they attempt to playout the chunk at buffer position $m$, causing a rightward shift of the buffer contents by one position. Since the data stream is real-time, peers are synchronized and they all attempt to playout the *same chunk* of data; in the figure the chunk labeled A is to be played out by all the peers. If the buffer position $m$ is empty, there is a gap in play out, and the missing chunk is never recovered (i.e., the rightward shift occurs even if the chunk is missing, and peers remain synchronized). Hence, the peer at the bottom of Figure 1 will never obtain chunk A.

Peers are part of a P2P network, and they can potentially obtain any chunk from each other until the point of playout of that chunk. Thus, chunks B through O can be obtained through P2P chunk sharing. Our P2P network model is a fully connected graph. Each peer chooses one peer at random, and may request a chunk from it. There are no restrictions on upload and download bandwidths, since with random peer selection the number of peers selecting any one peer is finite with high probability (so the results would essentially be the same even with such bandwidth constraints). Suppose that a peer has a set of chunks $\mathcal{Q}$ and its selected peer has a set of chunks $\mathcal{R}$. Define $\mathcal{A} = \mathcal{R} \setminus \mathcal{Q}$. Then a chunk selection *policy* $\mu$ is a rule by which the peer selects one of the chunks in $\mathcal{A}$.

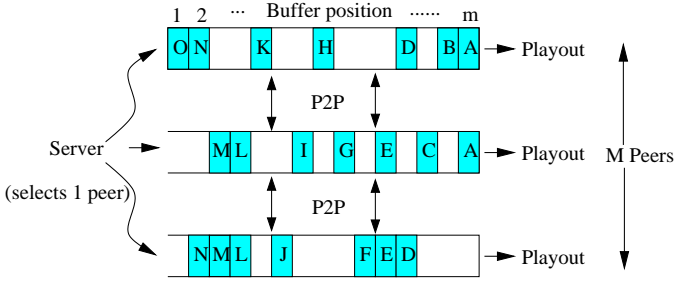

Fig. 1. A P2P content streaming system. Content is generated real-time at server, and each chunk is sent to one of $M$ interested peers. Peers try to obtain chunks via P2P among themselves. Chunks are shifted to the right by one position at each time step, and the chunk at position $m$ is played out if available. We would like to ensure that all peers possess the chunk at position $m$ with high probability.

The model in [19] is an equilibrium model, i.e., it studies the system assuming that it is in steady-state. We assume that the buffer occupancy probabilities have a steady-state distribution where the steady-state probability that buffer space $i$ is occupied is denoted $p_{(\mu,m)}(i)$. We assume that this distribution is identical and independent across peers. As we will see shortly, under the independence assumption, it is analytically tractable to characterize the impact of the rest of the system on a single peer. This is similar to mean-field approximations in physics where, in a large system, the impact of the rest of the system on a single particle is captured by a "mean field." In our case, the large-system assumption means that $M$ is assumed to be large.

Under the above assumption, we can write down a simple relation between the steady state probabilities (at the beginning of any time slot) of buffer occupancies as follows:

$$p_{(\mu,m)}(i+1) = p_{(\mu,m)}(i) + s_{(\mu,m)}(i) p_{(\mu,m)}(i)(1 - p_{(\mu,m)}(i)) \quad \forall\, i > 1, \quad (1)$$

$$p_{(\mu,m)}(1) = \frac{1}{M}. \quad (2)$$

In the above, since buffer position $i+1$ is filled by a rightward shift from buffer position $i$, its steady state probability at the beginning of the current time slot is the probability that $i$ was already filled at the beginning of the last time slot, plus the probability that $i$ was filled by P2P methods during the last time slot. The latter term is derived by considering that $p_{(\mu,m)}(i)(1 - p_{(\mu,m)}(i))$ is the probability that a peer does not possess $i$ but the selected peer does, and

$$s_{(\mu,m)}(i) = \sum_{\mathcal{A}: i \in \mathcal{A}} \mathbb{P}(\text{select } i | \mathcal{A}) \mathbb{P}(\mathcal{A}),$$

where $\mathcal{A}$ is the difference set between the peers as defined above. In other words, $s_{(\mu,m)}(i)$ is the probability that a peer $\pi$ chooses to download chunk $i$ from its selected peer $\lambda$, given that $\pi$ does not possess chunk $i$ while $\lambda$ does. This selection probability is a function of the chunk selection policy $\mu$. As mentioned earlier, a number of independence assumptions have been made in arriving at the above model; see [19] for an informal justification.

Two policies that are of particular interest are the following:

1) **Rarest-first:** Under this policy (denoted by $r$), priority is given to the chunks that have the lowest steady-state probability. From (1) for any policy $\mu$ we have $p_{(\mu,m)}(j) \leq p_{(\mu,m)}(i)$ for $j < i$. Hence, rarest-first is equivalent to selecting chunks with priority $2 > 3... > m-1$. An interesting result of [19] is that for this policy

$$s_{(r,m)}(i) = 1 - p_{(r,m)}(i). \quad (3)$$

2) **Greedy:** Under this policy (denoted by $g$), priority is given to the chunks that are closest to playout. From (1) this is equivalent to prioritizing those chunks that have the highest steady state probabilities. Hence, the greedy policy selects chunks with priority $m-1 > m-2... > 2$. It is shown in [19] that for the greedy policy

$$s_{(g,m)}(i) = 1 - \frac{1}{M} - p_{(g,m)}(m) + p_{(g,m)}(i+1). \quad (4)$$

The objective is to find a policy $\mu$ that has the smallest value of $m$ for a target value of $p_{(\mu,m)}(m)$. In other words, we would like to find a policy that requires the smallest buffer size for a desired probability of skip-free playout.

### III. Insights from a Fluid Approximation

Our first objective will be to obtain insight into the performance of the chunk-selection policies by approximating the difference equations in the previous section by differential equations. The resulting model is called the fluid model. While the fluid model is not precise, the main purpose of this model is to provide key intuition which will serve as the basis for our analysis in the next section, where will rigorously prove all results in the next section by directly working with the difference equations. Let us first consider the expression describing the steady state probabilities in (1). We can approximate this system of difference equations by using differential equations in the following manner.

$$\frac{dp_{(\mu,m)}(i)}{di} = s_{(\mu,m)}(i) p_{(\mu,m)}(i)(1 - p_{(\mu,m)}(i)), \quad (5)$$

$$p_{(\mu,m)}(0) = \frac{1}{M}. \quad (6)$$

Consistent with the fluid approximation, when applying the above equation to the greedy policy, we will replace $p_{i+1}$ by $p_i$ in the expression for $s_{(g,m)}(i)$ given in (4).

In this section, we will provide a heuristic explanation of the main ideas in the paper by using the above fluid model. In the later sections, we will use this intuition to derive our main results directly from the discrete model.

## A. Buffer sizing for the rarest-first policy

We first use the fluid model to study the minimum buffer size required to achieve a given probability of buffer occupancy. We have the following result.

*Fluid Result 1:* The buffer size $m$ required to attain $p_{(r,m)}(m) = q$ using the rarest-first policy is $\Theta(\log M) + \Theta(1/(1-q))$[1].

*Proof:* From (5) and (3), the dynamics of the rarest-first policy are given by
$$\frac{dp_{(r,m)}(i)}{di} = p_{(r,m)}(i)(1 - p_{(r,m)}(i))^2.$$

Solving the above differential equation using condition (6) yields
$$di = \frac{dp_{(r,m)}}{p_{(r,m)}} + \frac{dp_{(r,m)}}{1 - p_{(r,m)}} + \frac{dp_{(r,m)}}{(1 - p_{(r,m)})^2},$$
$$i = \ln\left(p_{(r,m)}(i)M\right) - \ln\left(\frac{1 - p_{(r,m)}(i)}{1 - 1/M}\right) + \frac{1}{1 - p_{(r,m)}(i)} - \frac{1}{1 - \frac{1}{M}}.$$

Thus, for a target $p_{(r,m)}(m) = q$, the required buffer size is given by
$$m = \ln(qM) - \ln\left(\frac{1-q}{1-1/M}\right) + \frac{1}{1-q} - \frac{1}{1-\frac{1}{M}}.$$

The desired result follows. ∎

## B. Buffer sizing for the greedy policy

We have the following upper and lower bounds on the minimum buffer requirements for the greedy algorithm.

*Fluid Result 2:* The buffer size $m$ required to attain $p_{(g,m)}(m) = q$ using the greedy policy is $O(\log M) + O((1/(1-q))\log(1/(1-q)))$.

*Proof:* From (4) that, with $p_{(g,m)}(m) = q$, we have
$$s_{(g,m)}(i) = 1 - q - \frac{1}{M} + p_{(g,m)}(i+1) \geq 1 - q, \quad (7)$$

where the inequality follows since $p_{(g,m)}(i)$ is increasing and $p_{(g,m)}(0) = 1/M$. Hence, the solution $p_{g,m}$ to
$$\frac{dp_{(g,m)}}{di} = s_{(g,m)}(i)p_{(g,m)}(i)(1 - p_{(g,m)}(i))$$
is greater than the solution $p_{g,m}$ to
$$\frac{dp_{(g,m)}}{di} = (1 - q)p_{(g,m)}(i)(1 - p_{(g,m)}(i)).$$

The latter equation can be solved as follows:
$$\frac{dp_{(g,m)}(i)}{di} = (1-q)p_{(g,m)}(i)(1 - p_{(g,m)}(i))$$
$$(1-q)di = \frac{dp_{(g,m)}(i)}{p_{(g,m)}(i)(1 - p_{(g,m)}(i))}$$
$$(1-q)i = \ln\frac{p(i)(1-p(0))}{(1-p(i))p(0)} \quad (8)$$

---
[1] We use the notation $\Theta(\log M) + \Theta(1/(1-q))$ to denote that, with $M$ fixed, the asymptote behaves like $\Theta(1/(1-q))$ and, with $q$ fixed, it behaves like $\Theta(\log M)$.

Using $p_{(g,m)}(0) = 1/M$, $p_{(g,m)}(m) = q$ and setting $i = m$, the above yields
$$m = \frac{1}{1-q}\ln\frac{q(1-1/M)}{(1-q)/M},$$
which gives the desired result. ∎

*Fluid Result 3:* The buffer size $m$ required to attain $p_{(g,m)}(m) = q$ using the greedy policy is $\Omega(\log M) + \Omega(1/(1-q))$.

*Proof:* Fix $\alpha \in (0, (q-1/M)/(1-q))$. Since $p_{(g,m)}(0) = 1/M$ and $p_{(g,m)}(m) = q$, there must be some $k \in (0, m)$ for which $p_{(g,m)}(k) = 1/M + \alpha(1-q)$. Then for all $i \leq k$, since $p(i)$ is an increasing function, we have the following upper bound on $s_{(g,m)}(i)$:
$$s(i) = 1 - q - \frac{1}{M} + p(i) \leq 1 - q - \frac{1}{M} + p(k) = c(q,M),$$
where
$$c(q,M) := (1+\alpha)(1-q).$$
Thus, the solution to
$$\frac{dp_{(g,m)}}{di} = s(i)p_{(g,m)}(i)(1 - p_{(g,m)}(i))$$
is upper bounded by the solution to
$$\frac{dp_{(g,m)}}{di} = c(q,M)p_{(g,m)}(i)(1 - p_{(g,m)}(i)).$$
Solving the second differential equation above yields
$$\ln\frac{p(i)(1-p(0))}{(1-p(i))p(0)} = c(q,M)i. \quad (9)$$
Letting $i = k$ and substituting $p(k) = 1/M + \alpha(1-q)$, gives
$$m \geq k = \frac{1}{c(q,M)}\ln\frac{1/M + \alpha(1-q)}{(1/M)} + \frac{1}{c(q,M)}\ln\frac{1 - 1/M}{1 - 1/M - \alpha(1-q)}.$$
It is easy to check from the above expression that $m$ is $\Omega(\log M)$ when $q$ is fixed and $\Omega(1/(1-q))$ when $M$ is fixed. ∎

## C. A hybrid policy

Our conclusions from the previous two subsections are:
- The buffer size requirement for both the rarest-first and greedy policies have logarithmic scaling in the number of users $\Theta(\log M)$.
- The buffer size requirement grows at least as $1/(1-q)$ for both policies when the desired skip-free playout probability is $q$.

This suggests that the buffer size requirement could be high if very stringent QoS is required, i.e., a skip-free playout probability close to $1$ is required. In the next sections, our goal is to understand if there is a different policy that can provide significant reduction in the buffer-size requirement. It has been observed in [19] that the buffer-size requirement can be reduced by a using a hybrid policy which uses the rarest-first policy under certain conditions and the greedy policy under certain conditions. Since both the rarest-first and greedy policies have similar asymptotic performance, we now use the fluid model to get some insight into when one policy performs better than the other, which would be helpful in designing and analyzing hybrid policies. From (3) we have that for the rarest-first policy

$$s_{(r,m)}(i) = 1 - p_{(r,m)}(i), \tag{10}$$

whereas for the greedy policy with a target $p_{(g,m)}(m) = q$, from (4)

$$\begin{aligned} s_{(g,m)}(i) &= p_{(g,m)}(i+1) + 1 - q - 1/M \\ &\approx p_{(g,m)}(i) + 1 - q - 1/M, \end{aligned} \tag{11}$$

where the approximation is motivated by the fluid model. Since $p_{(r,m)}(1) = p_{(g,m)}(1) = 1/M$, for $M > 2$ and small $(1-q)$, $s_{(g,m)}(1) < s_{(r,m)}(1)$. Thus, $p_{(r,m)}(2) > p_{(g,m)}(2)$. By induction, it follows that, for all $j$ such that $p_{(r,m)}(j) < 0.5$, we have $s_{(g,m)}(j) < s_{(r,m)}(j)$, which implies $p_{(r,m)}(j+1) > p_{(g,m)}(j+1)$. This suggests that, for the lower buffer positions, larger buffer occupancy probabilities can be obtained by using the rarest-first policy.

Next, let us consider the higher buffer positions, i.e., the ones numbered $m, m-1, \ldots$. Suppose that both the rarest-first and greedy policies exactly achieve $p_{(r,m)}(m) = p_{(g,m)}(m) = q$. Then, from (10) and (11), it follows that $s_{(g,m)}(m) > s_{(r,m)}(m)$. Due to the monotonicity of both $p_{(r,m)}(i)$ and $p_{(g,m)}(i)$, it follows that there exists a $k$ such that $s_{(g,m)}(i) > s_{(r,m)}(i)$ for all $i \geq m$. When $(1-q)$ is small and $M$ is large, such a $k$ would correspond to $p_{(g,m)}(k) = 0.5$. In the fluid model, since both $p_{(g,m)}(i)$ and $p_{(r,m)}(i)$ vary continuously as functions of $i$ (which is also a continuous variable in the fluid model), we make the following observation: to allow for the largest increase in buffer-occupancy probabilities, use the rarest-first policy till the buffer position where the occupancy probability is 0.5 and the then switch to the greedy policy. While this policy may not be optimal, the fluid model suggests that this may be a good heuristic to combine the rarest-first and greedy policies. In the next section, by directly working with the discrete-time model, we show that such a hybrid policy leads to significant reductions in the buffer size requirements for skip-free playout and is also optimal in an asymptotic order sense.

## IV. DISCRETE TIME MODEL: LOWER BOUNDS ON THE BUFFER SIZE REQUIREMENT

In this section, we obtain lower bounds on buffer size requirement. We first introduce two notations that will be extensively used in our analysis. We denote by $n_{(\mu,m);-q}$ the *largest* index $i$ of the buffer spaces such that $p_{(\mu,m)}(i) \leq q$, and $n_{(\mu,m);+q}$ the *smallest* index $i$ such that $p_{(\mu,m)}(i) \geq q$, i.e.,

$$\begin{aligned} n_{(\mu,m);-q} &= \max\left\{i : p_{(\mu,m)}(i) \leq q\right\} \\ n_{(\mu,m);+q} &= \min\left\{i : p_{(\mu,m)}(i) > q\right\}. \end{aligned}$$

It is easy to see that $n_{(\mu,m);-q} \leq n_{(\mu,m);+q} \leq n_{(\mu,m);-q} + 1$.

We first provide a simple lower bound on the buffer size, which holds for all chunk selection policies.

*Lemma 1:* For any $\frac{1}{M} \leq q \leq 1$, any chunk selection policy $\mu$ and any buffer size $m$, the following inequality holds

$$n_{(\mu,m);+q} \geq \log M + \log q.$$

*Proof:* First, we know that

$$\begin{aligned} & p_{(\mu,m)}(j+1) \\ &= p_{(\mu,m)}(j) + p_{(\mu,m)}(j)\left(1 - p_{(\mu,m)}(j)\right)s_{(\mu,m)}(i) \\ &= p_{(\mu,m)}(j)\left(1 + (1 - p_{(\mu,m)}(j))s_{(\mu,m)}(i)\right) \\ &\leq 2p_{(\mu,m)}(j). \end{aligned}$$

Since $p(0) = 1/M$, it follows from a simple induction argument that

$$p_{(\mu,m)}(j+1) \leq \frac{2^j}{M}.$$

Thus,

$$n_{(\mu,m);+q} \geq \log M + \log q.$$

∎

### A. Rarest-first policy

In this subsection, we obtain a lower bound on the buffer size required by the rarest-first policy.

*Theorem 2:* Given any target probability $0.5 < q \leq 1$, the buffer size required to achieve this target probability with the rarest-first policy is at least

$$\log M + \log(2q-1) + \frac{\log q - \log(2q-1)}{\log\left(1 + (2-2q)^2\right)} - 1.$$

*Proof:* Recall that $s_{(r,m)}(i) = 1 - p_{(r,m)}(i)$ and

$$p_{(r,m)}(j+1) = p_{(r,m)}(j) + p_{(r,m)}(j)\left(1 - p_{(r,m)}(j)\right)^2.$$

The above difference equation is difficult to solve, so we first lower-bound the number of buffer spaces to reach an occupancy probability of $(2q-1)$ and then further lower-bound the additional buffer space required to reach an occupancy probability of $q$.

First from Lemma 1, we have

$$n_{(r,m);+(2q-1)} \geq \log M + \log(2q-1), \tag{12}$$

which is a lower bound on the buffer size required to reach an occupancy probability $1 - 2(1-q) = 2q-1$. Next, we bound $n_{(r,m);-(2q-1)} - n_{(r,m);+q}$ (the additional buffer space required to increase the occupancy probability from $2q-1$ to $q$). Under the rarest-first policy, for any $j \geq n_{(r,m);+(2q-1)}$, we have

$$\begin{aligned} p_{(r,m)}(j+1) &= p_{(r,m)}(j) + p_{(r,m)}(j)\left(1 - p_{(r,m)}(j)\right)^2 \\ &\leq p_{(r,m)}(j)\left(1 + (2-2q)^2\right). \end{aligned}$$

Thus,

$$\begin{aligned} & p_{(r,m)}(n_{(r,m);+(2q-1)} + j) \\ &\leq p_{(r,m)}(n_{(r,m);+(2q-1)})\left(1 + (2-2q)^2\right)^j. \end{aligned} \tag{13}$$

To obtain an upper bound on $p_{(r,m)}(n_{(r,m);+(2q-1)})$, we know that

$$\begin{aligned}
& p_{(r,m)}(n_{(r,m);+(2q-1)}) \\
=& p_{(r,m)}(n_{(r,m);-(2q-1)}) + \\
& p_{(r,m)}(n_{(r,m);-(2q-1)}) \left(1 - p_{(r,m)}(n_{(r,m);-(2q-1)})\right)^2.
\end{aligned}$$

It is easy to verify that $x + x(1-x)^2$ is an increasing function in $x$ for $x \geq 0$, so we have

$$p_{(r,m)}(n_{(r,m);+(2q-1)}) \leq (2q-1)(1 + (2-2q)^2).$$

Substituting into inequality (13), we obtain

$$p_{(r,m)}(n_{(r,m);+(2q-1)} + j) \leq (2q-1) \left(1 + (2-2q)^2\right)^{j+1}.$$

Now we can conclude that

$$\begin{aligned}
q &\leq p_{(r,m)}(n_{(r,m);+q}) \\
&\leq (2q-1) \left(1 + (2-2q)^2\right)^{n_{(r,m);+q} - n_{(r,m);+(2q-1)} + 1},
\end{aligned}$$

which implies that

$$n_{(r,m);+q} - n_{(r,m);+(2q-1)} \geq \frac{\log q - \log(2q-1)}{\log\left(1 + (2-2q)^2\right)} - 1. \quad (14)$$

Lemma 1 is obtained by summing inequalities (12) and (14). ∎

**Remark:** Note that when $\delta$ is small enough, we have $\log(1+\delta) \approx \delta$. So

$$\frac{\log q - \log(2q-1)}{\log\left(1 + (2-2q)^2\right)} \approx \frac{-(1-q) + 2(1-q)}{4(1-q)^2} = \frac{1}{2(1-q)},$$

which implies that the buffer size required by the rarest first policy is $\Omega(\log M) + \Omega(\frac{1}{1-q})$.

### B. Greedy policy

In this subsection, we characterize the buffer size required by the greedy policy.

*Theorem 3:* Given any target probability $q \geq \frac{1}{M}$, the buffer size to attain this target probability with the greedy policy is at least

$$\log M + \log q - 1 + \frac{1}{\log\left(2 - q + \frac{2}{M}\right)}.$$

*Proof:* Recall that $s_{(g,m)}(i) = 1 - \frac{1}{M} - p_{(g,m)}(m) + p_{(g,m)}(i+1)$. Under the greedy policy,

$$\begin{aligned}
p_{(g,m)}(i+1) =& p_{(g,m)}(i) + p_{(g,m)}(i)(1 - p_{(g,m)}(i)) \times \\
& \left(1 - \frac{1}{M} - p_{(g,m)}(m) + p_{(g,m)}(i+1)\right).
\end{aligned}$$

Note that

$$p_{(g,m)}(n_{(g,m);+\frac{2}{M}}) \leq \frac{4}{M}$$

since $p_{(g,m)}(i+1) \leq 2p_{(g,m)}(i)$ holds for all $i$. Now if $p_{(g,m)}(m) \geq q$, then for any $i \leq n_{(g,m);-\frac{2}{M}}$, we have

$$\begin{aligned}
& p_{(g,m)}(i+1) \\
\leq& p_{(g,m)}(i) + p_{(g,m)}(i) \left(1 - \frac{1}{M}\right)\left(1 - q + \frac{3}{M}\right) \\
=& p_{(g,m)}(i) \left(1 + \left(1 - \frac{1}{M}\right)\left(1 - q + \frac{3}{M}\right)\right),
\end{aligned}$$

where the first inequality holds because

$$\begin{aligned}
p_{(g,m)}(i+1) &\leq p_{(g,m)}(n_{(g,m);-\frac{2}{M}} + 1) \\
&= p_{(g,m)}\left(n_{(g,m);+\frac{2}{M}}\right) \leq \frac{4}{M}.
\end{aligned}$$

Thus,

$$\begin{aligned}
& \frac{2}{M} \leq p_{(g,m)}(n_{(g,m);+\frac{2}{M}}) \\
\leq& p_{(g,m)}(1) \left(1 + \left(1 - \frac{1}{M}\right)\left(1 - q + \frac{3}{M}\right)\right)^{n_{(g,m);+\frac{2}{M}} - 1},
\end{aligned}$$

which yields

$$\begin{aligned}
n_{(g,m);+\frac{2}{M}} &\geq \frac{1}{\log\left(1 + \left(1 - \frac{1}{M}\right)\left(1 - q + \frac{3}{M}\right)\right)} + 1 \\
&\geq \frac{1}{\log\left(2 - q + \frac{2}{M}\right)} + 1.
\end{aligned}$$

Recall $p_{(g,m)}(i+1) \leq 2p_{(g,m)}(i)$ holds for all $i$. It is easy to verify that

$$n_{(g,m);+q} - n_{(g,m);-\frac{4}{M}} \geq \log M + \log q - 2.$$

Note that $n_{(g,m);-\frac{4}{M}} \geq n_{(g,m);+\frac{2}{M}}$ because $p_{(g,m)}(n_{(g,m);+\frac{2}{M}}) \leq \frac{4}{M}$, so we can conclude that

$$\begin{aligned}
n_{(g,m);+q} &\geq n_{(g,m);+q} - n_{(g,m);-\frac{4}{M}} + n_{(g,m);+\frac{2}{M}} \\
&\geq \log M + \log q - 1 + \frac{1}{\log\left(2 - q + \frac{2}{M}\right)}.
\end{aligned}$$
∎

**Remark:** When $1-q$ and $\frac{2}{M}$ are sufficiently small, we have

$$\frac{1}{\log\left(2 - q + \frac{2}{M}\right)} \approx \frac{1}{1 - q + \frac{2}{M}}.$$

According to the theorem above, when $1-q \geq \frac{1}{M}$, the buffer size needs to be at least $\log M + \frac{1}{2(1-q)}$; and when $1-q \leq \frac{1}{M}$, the buffer size needs to be at least $\log M + \frac{M}{4}$.

From Theorem 2 and Theorem 3, we can see that either the rarest-first policy or the greedy policy, if used along, require a buffer with size $\Omega(\log M) + \Omega(\frac{1}{1-q})$. Thus, if we have a target probability $p(n) = 99.9\%$, the buffer has to have at least 1000 chunk spaces. However, from the analysis, it is not difficult to see that under the rarest-first policy, the occupancy probability increases slowly when *close to $q$*; and under the greedy policy, the occupancy probability increases slowly *at the initial stage.* Similar observations can be found in heuristic argument and the simulations in [19]. This motivates us to consider hybrid policies that use the rarest-first policy on the buffer spaces with small indices and the greedy policy on the buffer spaces with large indices. We will demonstrate in the next section that a properly designed hybrid policy only requires $\Theta(\log M)$ buffer size for any target probability $q$ such that $q \geq 1 - \frac{1}{M}$.

### V. DISCRETE TIME MODEL: HYBRID POLICY

The insight obtained on hybrid policies from the fluid model of Section III was "use the rarest-first policy till the buffer position where the occupancy probability is 0.5 and the then switch to the greedy policy." In this section we will analyze this policy using a discrete model and show that this insight is indeed valid and achieves the minimum buffer size (in an order sense) required for a given miss probability. We first formally define the policy.

**Hybrid Policy:** Let $h_\epsilon$ denote a policy, where $\epsilon$ is the occupancy probability at which we switch from rarest-first to greedy. In other words, priority is given to buffer positions 1 to $n_{(r,\cdot);+\epsilon}$ with priority decreasing from $1 \to n_{(r,\cdot);+\epsilon}$ (rarest first). If the selected peer has no chunks in this set, then requests are made for chunks in position $n_{(r,\cdot);+\epsilon} + 1$ to $m$, with priority decreasing from $m \to n_{(r,\cdot);+\epsilon} + 1$ (greedy). The chunk selection function that determines the policy is then formally defined as follows:

- Considering any $i$ such that $i \leq n_{(r,\cdot);+\epsilon}$, the policy satisfies
$$S_{(h_\epsilon,m)}(\mathcal{A};i) = 1$$
if $i \in \mathcal{A}$ and $\mathcal{A} \cap \{1,\ldots,i-1\} = \emptyset$, and
$$S_{(h_\epsilon,m)}(\mathcal{A};i) = 0$$
otherwise.

- Considering any $i$ such that $i > n_{(r,\cdot);+\epsilon}$, the policy satisfies
$$S_{(h_\epsilon,m)}(\mathcal{A};i) = 1$$
if $i \in \mathcal{A}$ and $\mathcal{A} \cap \{1,\ldots,n_{(r,m);+\epsilon)}, i+1,\ldots,m\} = \emptyset$, and
$$S_{(h_\epsilon,m)}(\mathcal{A};i) = 0$$
otherwise.

We now analyze this policy and show that the buffer size requirement of this policy is optimal in the order sense.

*Theorem 4:* Under the hybrid policy $h_\epsilon$, if we have a target occupancy of the final buffer space $q \geq 0.8$, and the switching parameter $\epsilon = 0.5$, if
$$m \geq \frac{1}{\log \frac{1}{1-\delta(1-\epsilon_2-q-\delta)}} \log \frac{1-\delta}{1-q} + 2$$
$$+ \frac{1}{\log \frac{1}{1-\alpha}} \log \frac{\delta}{1-q} + \frac{\log 2M\epsilon}{\log(1+(1-\epsilon)^2)},$$
then $p_{h_\epsilon,m}(m) \geq q$. Here, $\delta = 0.8$; $\alpha = \epsilon(1-\delta)$; $\epsilon_2 = \epsilon_1(1+(1-\epsilon_1)^2)$; $\epsilon_1 = \epsilon(1+(1-\epsilon)^2)$.

*Proof:* The proof is in three parts. We make the supposition that $p_{h_\epsilon,m}(m) < q$, and prove that it results in a lower bound on the buffer size $m$ that violates our assumptions.

**Step 1.** In the first part, we calculate the buffer size required for the rarest first policy to reach an occupancy probability that exceeds $\epsilon > 0$. Formally, we will determine the value of $n_{(r,\cdot);+\epsilon}$, which for simplicity of notation within the proof, we will denote by $n_{+\epsilon}$. Recall from (1) and (3) that since we use the rarest first policy for buffer positions $i \leq n_{+\epsilon} - 1$, the occupancy probability of buffer position $i$ is described by

$$\begin{aligned} p_{(r,\cdot)}(i+1) &= p_{(r,\cdot)}(i)\left(1 + (1-p_{(r,\cdot)}(i))^2\right) \quad (15) \\ &\geq p_{(r,\cdot)}(i)\left(1 + (1-\epsilon)^2\right) \\ &\geq p(1)\left(1+(1-\epsilon)^2\right)^i, \end{aligned}$$

where we note that $p(1) = 1/M$ is independent of the policy used. Thus,

$$\begin{aligned} p_{(r,\cdot)}(n_{+\epsilon}) &\geq p(1)\left(1+(1-\epsilon)^2\right)^{n_{+\epsilon}} \\ \Rightarrow n_{+\epsilon} &\leq \frac{\log M p_{(r,\cdot)}(n_{+\epsilon})}{\log\left(1+(1-\epsilon)^2\right)}. \end{aligned} \quad (16)$$

Also we have from (15) that

$$p_{(r,\cdot)}(n_{+\epsilon}) = $$
$$p_{(r,\cdot)}(n_{+\epsilon}-1)\left(1+\left(1-p_{(r,\cdot)}(n_{+\epsilon}-1)\right)^2\right)$$
$$\leq \epsilon\left(1+(1-\epsilon)^2\right) \leq 2\epsilon, \quad (17)$$

where the last line follows from the fact that $x(1+(1-x)^2)$ is an increasing function. Finally, from (16) and (17), we have

$$n_{+\epsilon} \leq \frac{\log 2M\epsilon}{\log\left(1+(1-\epsilon)^2\right)}. \quad (18)$$

We next characterize the performance of the greedy part of the hybrid policy.

**Step 2.** We first determine the buffer space required for the hybrid policy to exceed an occupancy probability $\delta > \epsilon$. In other words, we calculate $n_{(h_\epsilon,m);+\delta}$, and similarly characterize $n_{(h_\epsilon,m);-\delta} = n_{(h_\epsilon,m);+\delta} - 1$. Again, for simplicity of notation within the proof, these will be denoted as $n_{+\delta}$ and $n_{-\delta}$, respectively. Consider the set of buffer positions $j$ that satisfy $n_{+\epsilon} + 1 \leq j \leq n_{-\delta}$. From Lemma 5 (see Appendix) and using $a \triangleq p_{(h_\epsilon,m)}(n_{+\epsilon}+1)$ for convenience, it is easy to see that

$$\begin{aligned} p_{(h_\epsilon,m)}(j+1) &= \\ p_{(h_\epsilon,m)}(j) &+ p_{(h_\epsilon,m)}(j)(1-p_{(h_\epsilon,m)}(j)) \\ &\times \left(1-a-p_{(h_\epsilon,m)}(m)+p_{(h_\epsilon,m)}(j+1)\right) \quad (19)\\ &\geq p_{(h_\epsilon,m)}(j) + \\ &\epsilon(1-\delta)(1-a-p_{(h_\epsilon,m)}(m)+p_{(h_\epsilon,m)}(j+1)). \end{aligned}$$

Now, choosing $p_{(h_\epsilon,m)}(m) < q$, we have

$$p_{(h_\epsilon,m)}(j+1) \geq$$
$$p_{(h_\epsilon,m)}(j) + \epsilon(1-\delta)(1-a-q+p_{(h_\epsilon,m)}(j+1))$$
$$\Rightarrow (1-\alpha)p_{(h_\epsilon,m)}(j+1) \geq p_{(h_\epsilon,m)}(j) + \alpha(1-a-q),$$

where $\alpha \triangleq \epsilon(1-\delta)$. The above inequality allows us to recursively calculate $p_{(h_\epsilon,m)}(n_{+\epsilon}+k)$ for $k \geq 1$, and we will use it in order to determine $n_{-\delta}$. We have

$$(1-\alpha)p_{(h_\epsilon,m)}(n_{+\epsilon}+2) \geq p_{(h_\epsilon,m)}(n_{+\epsilon}+1) + \alpha(1-a-q),$$
$$(1-\alpha)p_{(h_\epsilon,m)}(n_{+\epsilon}+3) \geq p_{(h_\epsilon,m)}(n_{+\epsilon}+2) + \alpha(1-a-q).$$

Hence we have

$$(1-\alpha)^2 p_{(h_\epsilon,m)}(n_{+\epsilon}+3) \geq $$
$$(1-\alpha)p_{(h_\epsilon,m)}(n_{+\epsilon}+2) + \alpha(1-\alpha)(1-a-q),$$
$$\geq p_{(h_\epsilon,m)}(n_{+\epsilon}+1) + \alpha(1-a-q) + \alpha(1-\alpha)(1-a-q). \quad (20)$$

We can generalize the above as

$$(1-\alpha)^{k-1} p_{(h_\epsilon,m)}(n_{+\epsilon}+k)$$
$$\geq p_{(h_\epsilon,m)}(n_{+\epsilon}+1) + \alpha(1-a-q)\sum_{i=1}^{k-2}\alpha(1-\alpha)^i$$
$$= a + (1-a-q)(1-\alpha)\left(1-(1-\alpha)^{k-2}\right),$$

where we have used our definition $a \triangleq p_{(h_\epsilon,m)}(n_{+\epsilon}+1)$. Choosing $q > \epsilon = 0.5$ implies that $a + q > 1$, which in turn means that $1-a-q < 0$. Then since $(1-\alpha)(1-(1-\alpha)^{k-2}) < 1$, we have the relation

$$(1-\alpha)(1-a-q)(1-(1-\alpha)^{k-2}) \geq 1-a-q. \quad (21)$$

From (20) and (21) we then have

$$(1-\alpha)^{k-1} p_{(h_\epsilon,m)}(n_{+\epsilon}+k) \geq a + (1-a-q) = 1-q. \quad (22)$$

We are now in a position to obtain a bound on $n_{-\delta}$. Let $k$ be such that $n_{+\epsilon}+k = n_{-\delta}$. Also, by definition, $p_{(h_\epsilon,m)}(n_{-\delta}) \leq \delta$. Using this fact, we see from (22) that

$$\delta(1-\alpha)^{k-1} \geq 1-q$$
$$\Rightarrow k \leq 1 + \frac{1}{\log(1-\alpha)}\log\frac{1-q}{\delta}.$$

In summary, we have now established that

$$n_{-\delta} - n_{+\epsilon} \leq 1 + \frac{1}{\log\frac{1}{(1-\alpha)}}\log\frac{\delta}{1-q}, \quad (23)$$

where $\alpha = \epsilon(1-\delta)$ and we chose $\epsilon = 0.5$.

**Step 3.** Our final step is to characterize the buffer size required to take the probability of occupancy from $\delta$ to our target $p_{h_\epsilon,m}(m) \geq q$. We first obtain an upper bound on $p_{h_\epsilon,m}(n_{+\epsilon}+1)$ that will be used in our analysis. Recall that the rarest-first policy is used up to and including buffer position $n_{+\epsilon}$. This means that from (15) we have

$$p_{(h_\epsilon,m)}(n_{+\epsilon}+1) = p_{(r,\cdot)}(n_{+\epsilon})\left(1 + (1-p_{(r,\cdot)}(n_{+\epsilon}))^2\right)$$

Then since $p_{(r,\cdot)}(n_{-\epsilon}) \leq \epsilon$ and $x(1+(1-x)^2)$ is an increasing function

$$p_{(r,\cdot)}(n_{+\epsilon}) \leq \epsilon(1+(1-\epsilon)^2) \triangleq \epsilon_1 \quad (24)$$
$$p_{(h_\epsilon,m)}(n_{+\epsilon}+1) \leq \epsilon_1(1+(1-\epsilon_1)^2) \triangleq \epsilon_2, \quad (25)$$

and we have chose $\epsilon = 0.5$. We now consider the evolution of occupancy probability from $\delta \to q$. From (19) we have

$$1 - p_{(h_\epsilon,m)}(j+1) =$$
$$1 - p_{(h_\epsilon,m)}(j) - p_{(h_\epsilon,m)}(j)(1-p_{(h_\epsilon,m)}(j))$$
$$\times \left(1 - a - p_{(h_\epsilon,m)}(m) + p_{(h_\epsilon,m)}(j+1)\right)$$
$$\leq (1-p_{(h_\epsilon,m)}(j))\left(1 - \delta\left(1 - \epsilon_2 - p_{(h_\epsilon,m)}(m) + \delta\right)\right), \quad (26)$$

where we have used $a \triangleq p_{(h_\epsilon,m)}(n_{+\epsilon}+1) \leq \epsilon_2$. Now, suppose that $p_{(h_\epsilon,m)}(m) < q$. Then from (26) we have

$$1 - p_{(h_\epsilon,m)}(j+1) < (1-p_{(h_\epsilon,m)}(j))\left(1 - \delta(1-\epsilon_2-q+\delta)\right).$$

In particular, by recursion on the above starting at $n_{+\delta}$

$$1 - q < 1 - p_{(h_\epsilon,m)}(m)$$
$$< (1-p_{(h_\epsilon,m)}(n_{+\delta}))\left(1-\delta(1-\epsilon_2-q+\delta)\right)^{m-n_{+\delta}}$$
$$< (1-\delta)\left(1-\delta(1-\epsilon_2-q+\delta)\right)^{m-n_{+\delta}}$$

We can rewrite the above in the following manner

$$(m-n_{+\delta})\log\left(1-\delta(1-\epsilon_2-q+\delta)\right) > \log\frac{1-q}{1-\delta}.$$

Now, by assumption $\delta < q$ so the right side of the above is negative. If we can find $0 < \delta < q$ for which the left side is negative, we can obtain an upper bound on $m$. In other words, we search for $\delta$ that satisfies

$$0 < (1-\epsilon_2-q+\delta) < 1$$
$$\Rightarrow q+\epsilon_2-1 < \delta < q+\epsilon_2, \quad (27)$$

where from (25) we have $\epsilon_2 < 0.72$. Also, since $\epsilon = 0.5$, we have $\epsilon + q > 1$. Thus, sufficient conditions on $\delta$ are

$$\delta \geq q - 0.28 > q+\epsilon_2-1 \text{ and}$$
$$\delta < 1 < q+\epsilon_2.$$

For example, we could choose $\delta = 0.8$, with $q \geq 0.8$. Thus,

$$m \leq n_{+\delta} + \frac{1}{\log\frac{1}{1-\delta(1-\epsilon_2-q+\delta)}}\log\frac{1-\delta}{1-q}. \quad (28)$$

Finally, from the three bounds (18), (23) and (28) we have a contradiction of the assumption on the size of buffer. Hence, we must have $p_{(h_\epsilon,m)}(m) \geq q$, which yields the proof. ■

## VI. SIMULATIONS

In this section, we use simulations to further evaluate the performance of different chunk-selection policies. We first consider a network with a fixed number of active peers. Specifically, the network consists of one server and 10,000 peers. Each peer has a buffer of size $m$. The network is a slotted time system. During each time slot, the server randomly selects a peer and uploads a new chunk to it, and each peer (expect the one who obtains the new chunk from the server) randomly selects another peer and downloads a chunk selected according to the chunk-selection policy. Figure 2 shows the minimum buffer sizes for attaining target skip-free playout probabilities under the greedy, rarest-first and hybrid polices. We see that the buffer size required under the hybrid policy is substantially smaller than the rarest-first and greedy policies. According to our simulation, the greedy policy requires the buffer size to be 183 to attain skip-free playout probability 0.976, the rarest-first requires the buffer-size to be 166 to attain skip-free playout probability 0.996, and *the hybrid policy only requires a buffer size of* 40 *to attain skip-free playout probability* 0.999!

In reality, the number of active peers in a P2P network changes over time because peers can dynamically join and leave. To study the performance of different polices with peer arrivals/departures, we consider a network with $20,000$ peers. Initially, $10,000$ peers are active. At each iteration, an active peer becomes inactive with probability $0.001$, and an inactive peer becomes active with probability $0.001$.

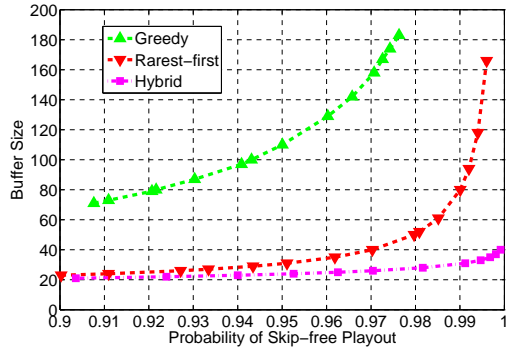

Fig. 2. The minimum buffer size versus target skip-free playout probability with a fixed number of peers in the system. The hybrid policy performance is an order better than either greedy or rarest first.

In our simulation, a peer empties its buffer when it is inactive, and begins to play the video $m$ time slots (start-up latency) after it becomes active, where $m$ is the buffer size. Figure 3 shows the minimum buffer sizes for attaining target skip-free playout probabilities under the rarest-first and hybrid polices. The skip-free probability is computed based on peers who are playing the video (not including peers who are in their start-up phase). The greedy policy is not included because its performance is really poor; for example, the skip-free playout probability of the greedy policy with buffer size $200$ is still less than $0.90$. From Figure 3, we see that the buffer size required under the hybrid policy is much smaller than the rarest-first policy. In fact, the rarest-first requires the buffer-size to be $125$ to attain skip-free playout probability $0.99$, whereas *the hybrid policy only requires the buffer size to be* $39$*!* This simulation indicates that the hybrid policy works well even in P2P networks with peer arrivals/departures.

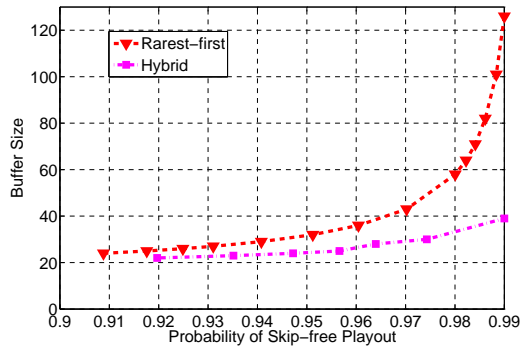

Fig. 3. The minimum buffer size versus skip-free playout probability with peer arrivals and departures. Greedy is not shown due its extremely poor performance. The hybrid policy easily does an order better than rarest-first.

## VII. CONCLUSION

In this paper we considered the problem of designing efficient policies for real-time streaming applications using P2P approaches. Our objective was to ensure that the playout buffer for a given target of skip free playout remains as small as possible. We showed that both the rarest-first and greedy policy have similar buffer scalings, and that their combination into a hybrid policy yielded order sense improvements in the required buffer size. Further, the buffer size required by the hybrid policy is close to the minimum over all policies. Future work includes the design of policies which are designed for ensuring good QoS for specific video codec formats.

## APPENDIX

*Lemma 5:* Consider the hybrid policy $h_\epsilon$. For buffer space $i$ such that $i \geq n_{(r,\cdot),+\epsilon_1}$, the steady state probabilities of occupancies satisfy

$$p_{(h_\epsilon,m)}(j+1) = \\ p_{(h_\epsilon,m)}(j) \; + \; p_{(h_\epsilon,m)}(j)(1 - p_{(h_\epsilon,m)}(j)) \\ \times \left(1 - a - p_{(h_\epsilon,m)}(m) + p_{(h_\epsilon,m)}(j+1)\right),$$

where $a \triangleq p_{(r,\cdot)}(n_{(r,\cdot),+\epsilon} + 1)$.

*Proof:* Recall that the hybrid policy first selects a threshold buffer position where the steady state occupancy probability using the rarest-first policy is greater than $\epsilon$ (this position is called $n_{(r,\cdot),+\epsilon}$), gives priority to the rarest-first algorithm, and uses the greedy algorithm if the difference set with the selected peer contains none of the chunks indexed from 1 to $n_{(r,\cdot),+\epsilon}$. We first find the probability that the greedy algorithm is used. Let $\gamma = 1 - \frac{1}{M}$. According to the definition of the hybrid policy, we have for $n_{(r,\cdot),+\epsilon} \geq 1$,

$$\gamma \tilde{p}_{(h_\epsilon,m)} \left( \mathcal{A} : \mathcal{A} \bigcap \{1, \ldots, n_{(r,\cdot),+\epsilon}\} = \emptyset \right)$$
$$= \gamma \prod_{j=1}^{n_{(r,\cdot),+\epsilon}} \left(1 - p_{(h_\epsilon,m)}(j)\left(1 - p_{(h_\epsilon,m)}(j)\right)\right)$$
$$= \gamma \prod_{j=1}^{n_{(r,\cdot),+\epsilon}} \left(1 - p_{(r,m)}(j)\left(1 - p_{(r,m)}(j)\right)\right)$$
$$= s_{(r,\cdot)}(n_{(r,\cdot),+\epsilon} + 1)$$
$$= \left(1 - p_{(r,m)}(n_{(r,\cdot),+\epsilon} + 1)\right)$$
$$\triangleq 1 - a.$$

So $1-a$ is the probability that none of the chunks indexed from 1 to $n_{(r,\cdot),+\epsilon}$ is in the difference set. When this event happens, the hybrid policy uses the greedy policy, and the proof next is similar to the proof of Proposition 1 of [19]. Following (1) and (4), the steady-state probabilities of occupancies can be written as

$$p_{(h_\epsilon,m)}(i+1)$$
$$= p_{(h_\epsilon,m)}(i) + p_{(h_\epsilon,m)}(i)(1 - p_{(h_\epsilon,m)}(i)) \times$$
$$(1-a) \prod_{j=i+1}^{m-1} \left(1 - p_{(h_\epsilon,m)}(j)\left(1 - p_{(h_\epsilon,m)}(j)\right)\right).$$

Defining $s(i) = \prod_{j=i+1}^{m-1} \left(1 - p_{(h_\epsilon,m)}(j)\left(1 - p_{(h_\epsilon,m)}(j)\right)\right)$ for $i < m-1$ and $s(m-1) = 1$, the equation above can be written as

$$p_{(h_\epsilon,m)}(i+1) =$$
$$p_{(h_\epsilon,m)}(i) + (1-a)s(i)p_{(h_\epsilon,m)}(i)(1 - p_{(h_\epsilon,m)}(i)) \quad (29)$$

and

$$\frac{s(i)}{s(i+1)} = 1 - p_{(h_\epsilon,m)}(i+1)(1 - p_{(h_\epsilon,m)}(i+1)),$$

which implies that

$$s(i+1) - s(i) = s(i+1)p_{(h_\epsilon,m)}(i+1)(1 - p_{(h_\epsilon,m)}(i+1)).$$

Substituting the equality above into equality (29), we obtain

$$p_{(h_\epsilon,m)}(i+2) - p_{(h_\epsilon,m)}(i+1) = (1-a)(s(i+1) - s(i)).$$

Summing up the inequalities above from $i$ to $m-2$, we have

$$p_{(h_\epsilon,m)}(m) - p_{(h_\epsilon,m)}(i+1) = (1-a)(s(m-1) - s(i)).$$

Note that $s(m-1) = 1$, so

$$(1-a)s(i) = 1 - a - p_{(h_\epsilon,m)}(m) + p_{(h_\epsilon,m)}(i+1),$$

and the lemma holds. ∎